\documentclass[aps,prl,twocolumn]{revtex4}
\usepackage{amsmath}
\usepackage{graphicx}
\begin{document}

\title{Intrinsic Domain Wall Resistance in Ferromagnetic Semiconductors}

\author{Anh Kiet Nguyen, R. V. Shchelushkin, and Arne Brataas}

\affiliation{Department of Physics, Norwegian University of Science and 
Technology, N-7491, Trondheim, Norway}

\begin{abstract}
  Transport through zincblende magnetic semiconductors with magnetic
  domain walls is studied theoretically. We show that these magnetic
  domain walls have an intrinsic resistance due to the effective hole
  spin-orbit interaction. The intrinsic resistance is independent of
  the domain wall shape and width, and survives the adiabatic limit.
  For typical parameters, the intrinsic domain wall resistance is
  comparable to the Sharvin resistance and should be experimentally
  measurable.
\end{abstract}

\maketitle

\newcommand{\eq}{\! = \!}


Understanding magnetic topological defects is crucial in developing
devices that utilize the electron spin. Domain walls are topological
defects between homogeneous magnetic domains. The domain wall dynamics
has traditionally been induced by external magnetic fields. There has
recently been a large interest in the scientific community on current
induced magnetization dynamics, where domain walls and domains change
in response to applied electric currents \cite{cims}. Domain walls can
also be electrically detected, by their electrical resistance.
Knowledge of the domain wall's effect on the electrical resistance is
important for the understanding of spin transport in condensed matter
and for the electrical detection of magnetic topological defects.

Transport through domain walls have been extensively studied in
metallic systems, theoretically \cite{dwth} and experimentally
\cite{dwexp}.  The domain wall resistance is defined as $R_{w} \eq R
\! - \! R_0$, where $R$ and $R_0$ are resistances with a domain wall
and with homogeneous magnetization, respectively. When the domain wall
is thinner than the mean free path, in the ballistic regime, $R_{w}$
is positive. In diffusive systems, when the domain wall is wider than
the mean free path, the sign of the domain wall resistance is still
under debate, \textit{i.e.} the domain wall can increase or decrease
the resistance of the ferromagnet. In ballistic and diffusive metal
systems, $R_{w}$ approaches zero with increasing domain wall width and
vanishes in the adiabatic limit when the domain wall is much wider
than the Fermi wavelength.

Ferromagnetic semiconductors integrate magnetization controlled spin
transport with gate controlled carrier densities in semiconductors.
Domain walls in these systems have been recently studied
experimentally
\cite{Ruster:prl03,Yamanouchi:Nature04,Tang:Nature04,Chiba:prl06}.
The strong interaction between the spin of the effective holes and
their orbits in dilute magnetic semiconductors changes the transport
properties of magnetic domain walls qualitatively.  In this Letter, we
show that domain walls in zincblende magnetic semiconductors have an
intrinsic resistance $R_{w}^I$ which is the part of $R_w$ that
survives the adiabatic limit.  $R_{w}^I$ is independent of the width
and detailed shape of the domain walls and is due to the effective
spin-orbit interaction.

Related manifestations of the coupling between the spin-orbit
interactions and the magnetizations are the anisotropic magneto
resistance (AMR)
\cite{Hayashi:physicaB00,Baxter:prb02,Jungwirth:apl03,Wang:prb05} and
the tunneling anisotropic magneto resistance (TAMR)
\cite{Gould:prl04,Ruster:prl03,Brey:apl04}. In domain walls, some
carriers are prevented by the spin-orbit interaction to adiabatically
adapt to the change in the direction of the local magnetization
producing a finite $R_w$ even in the adiabatic limit. $R^I_{w}$
depends mainly on three material parameters: the effective spin-orbit
coupling, the exchange field and the Fermi energy.  Hole transport in
dilute magnetic semiconductors with strong spin-orbit interaction is
usually difficult to treat analytically.  Nevertheless, we find an
{\ em analytical solution for the adiabatic domain wall conductance}.

We model hole transport in zincblende magnetic semiconductors by the 
following dimensionless Hamiltonian
\begin{equation}
  H =   \alpha_1 p_i p_i - \alpha_2 (J_i J_j p_i p_j) 
      + J_i h_i, \label{Hamiltonian}
\end{equation}
where the subscripts $(i,j) \eq x,y,z$ and the Einstein sum convention
is assumed.  Furthermore, $\mathbf{h}(\mathbf{r})$ is the
dimensionless exchange field, describing the interaction between holes
and localized magnetic moments, $|\mathbf{h}(\mathbf{r})| \eq h_0$.
In Eq.\ \ref{Hamiltonian}, $\mathbf{p}$ is the dimensionless momentum
operator and $\mathbf{J}$ denotes the angular momentum operator for $J
\eq 3/2$ spins. The parameters $\alpha_1$ and $\alpha_2$ are
controlled by the hole effective mass and the strength of the
spin-orbit interaction, respectively. Their relation to the Luttinger
parameters \cite{Luttinger:pr56} $\gamma_1$, $\gamma_2$ and $\gamma_3$
are $\alpha_1 \eq (\gamma_1 \!+\! 5\gamma_2/2)/(\gamma_1 \!-\! 2
\gamma_2)$ and $\alpha_2 \eq 2 \gamma_2 / (\gamma_1 \!-\! 2
\gamma_2)$. In our notation, energies, momentums and lengths are
measured in units of the Fermi energy, Fermi momentum and Fermi
wavelength of heavy holes for a given doping in the original
non-magnetic host system, $h_0 \eq 0$.  A six band version of Eq.\
\ref{Hamiltonian} explains many features of zincblende magnetic
semiconductors \cite{Dietl:Science00,Abolfath:prb01,Konig:book02}. We
employ the spherical approximation \cite{Baldereschi:prb73} and
disregard two split-off bands. Thus, $\gamma_3 \eq \gamma_2$ and the
Fermi energy $E_F$ is assumed to be smaller than the split-off energy.
We compute the conductance $G \eq 1/R$ from the Landauer-B\"{u}ttiker
formula $G \eq (e^2/h) \sum_{nm} |t_{nm}|^2$, where $t_{nm}$ is the
unit flux normalized transmission between the $n$ and $m$ transverse
waveguide modes.

Let us first prove that holes governed by the Hamiltonian
(\ref{Hamiltonian}) exhibit an intrinsic adiabatic domain wall
resistance.  We separate the Hamiltonian into an intrinsic
contribution $\tilde{H}_I$ and a collision contribution $\tilde{H}_C$
by transforming $H$ into a local basis where the quantization axis for
$\mathbf{J}$ is parallel to $\mathbf{h}(\mathbf{r})$, $\tilde{H} \eq U H
U^{-1} \eq \tilde{H}_I \!+\! \tilde{H}_C$:
\begin{eqnarray}
  \tilde{H}_I = \alpha_1 p_i p_i - 
              \alpha_2 ( \tilde{J}_i\tilde{J}_j p_i p_j) + 
              J_z h_0  \, , 
\label{Hamiltonian:Rotated} \\
  \tilde{H}_C = 
        (\alpha_1 \delta_{ij} - \alpha_2 \tilde{J_i} \tilde{J_j}) 
        [ \kappa_i  \kappa_j + \Lambda_{ij} + 2 \kappa_i p_j ] \, ,
\nonumber
\end{eqnarray}
where $U(\mathbf{r})$ is a $4 \times 4$ unitary operator defined such
that $U [\mathbf{h}(\mathbf{r}) \cdot \mathbf{J}] U^{-1} \eq J_z h_0$.
Furthermore, $\tilde{J}_i \eq U J_i U^{-1}$, $\kappa_i \eq U p_i U^{-1}$
and $\Lambda_{ij} \eq p_i U p_j U^{-1}$ are $4 \times 4$ matrices that
do not operate on the spatial coordinate $\mathbf{r}$.
\\
Consider the adiabatic limit when the width of the domain wall
$\lambda_{w}$ is much larger than the Fermi wavelength $\lambda_F$.
Hence, $\kappa_i  \!\sim\! 1/\lambda_{w} \rightarrow 0$, $\Lambda_{ij}  \!\sim\!
1/\lambda_{w}^2 \rightarrow 0$ and thus $\tilde{H}_C \rightarrow 0$.
Without the effective spin-orbit coupling, $\alpha_2 \eq 0$,
$\tilde{H}_I$ is independent of directional variations in the exchange
field and the domain wall resistance vanishes, $R_{w} \eq 0$. In
contrast, with finite effective spin-orbit coupling, $\alpha_2 \not =
0$, $\tilde{H}_I$ varies for systems with or without variations in the
exchange field, since $\tilde{J}_i$ differs from $J_i$. In other
words, the spin polarization of the transport channels is anisotropic,
giving rise to a finite intrinsic domain wall resistance.

In order to quantify our findings, we consider the linear response of
a rectangular conductor with a cross section $A \eq L_x L_z$ and
periodic boundary condition (PBC) along the $x$ and $z$ directions.
Transport is along the $y$-axis. We consider three types of domain
walls: Bloch ZX-wall, Neel ZY wall and Neel YZ-wall described by
$\mathbf{h}(y) \eq [f_2(\tilde{y}),0,f_1(\tilde{y})]$, $\mathbf{h}(y)  \eq 
[0,f_2(\tilde{y}),f_1(\tilde{y})]$ and $\mathbf{h}(y)  \eq 
[0,f_1(\tilde{y}),f_2(\tilde{y})]$, respectively.  Here,
$f_1(\tilde{y}) \eq h_0 \tanh(y/\lambda_{w})$ and $f_2(\tilde{y}) \eq
h_0/\cosh(y/\lambda_{w})$.

Translation invariance conserves the transverse momenta $k_x$ and
$k_z$ that label the transport channels. Additionally, each
($k_x,k_z$) channel contains four internal spin channels originating
from the four spin-orbit coupled bands.  The distribution of the
transport channels at position $y$, $T_y(k_x,k_z)$, may be found by
solving the eigenvalue problem $H(k_x,y,k_z) \psi_y \eq E_F \psi_y$.
The Fermi energy, $E_F$, is position independent.  Due to the
interplay of the spin-orbit interaction and the magnetization, the
number of open channels is anisotropic in the momentum $k_x$-$k_z$
space, unlike systems where the effective spin-orbit interaction
vanishes. In the adiabatic limit, a conducting channel must exists
throughout the system in order to contribute to the conductance.
Mathematically this can be expressed as $G \eq (e^2/h) \sum_{k_x,k_z}
T(k_x,k_y)$, where $T(k_x,k_y) \eq \cap_{y=-\infty}^{\infty}
T_y(k_x,k_z)$ is the intersection of all distributions of transport
channels as one traverses the domain wall, see Fig.1. In the case of a
homogeneous ferromagnet, the conductance is $G_0 \eq 1/R_0 \eq (e^2/h)
\sum_{k_x,k_z} T_{y=-\infty}(k_x,k_z)$.  If and only if
$T_{y=-\infty}$ is identical to or is a subset of $T_y$ for all $y$
then $R^I_{w} \eq 0$.  This occurs only when the effective spin-orbit
interaction or $h_0$ vanishes.

For Bloch walls, $T_y(k_x,k_z)$ is conformal under translation and
simply rotates in the same direction as the exchange field, see Fig.1.
Traversing from $y \eq -\infty$ to $\infty$ corresponds to a
$\pi$-rotation of $T_{y=-\infty}(k_x,k_z)$ forcing $T(k_x,k_y)$ to be
circular symmetric, see Fig.1. An analytical expression for the
ballistic conductance through an adiabatic Bloch wall can be written
as
\begin{eqnarray}
  G_{B} = \frac{e^2 A }{2 \hbar} \left \{ \Re\mathbf{e}(k_1)^2 + 
           \Re\mathbf{e}(k_2)^2 + 
           \Re\mathbf{e}(k_3)^2 + \Re\mathbf{e}(k_4)^2 \right \} 
\label{Conductance:BW}
\end{eqnarray}
where $\Re\mathbf{e}$ denotes the real part and
\begin{eqnarray}
  k_1 &=& \sqrt{(2 \gamma_1 E_F + D)B} ~~~,
  k_4 = \sqrt{(2 \gamma_1 E_F - D)B}
\nonumber \\ 
  k_2 &=& \Xi(X_1) \sqrt{E_F -3h_0/2}
\nonumber \\
      &+& \Xi(-X_1)
          \sqrt{(\gamma_1-2\gamma_2)(2E_F + h_0)B}
\nonumber \\ 
  k_3 &=& \Xi(X_1)
          \sqrt{(\gamma_1-2\gamma_2)(2 E_F+h_0)B}
\nonumber \\ 
      &+& \Xi(-X_1) ~ \Xi(X_2) \sqrt{E_F - 3 h_0/2}
\nonumber \\ 
      &+& \Xi(-X_2)
          \sqrt{(\gamma_1-2\gamma_2)(2E_F - h_0)B}
\nonumber
\end{eqnarray}
where $\Xi(X)$ is the Heaviside step function with $\Xi(0) \eq 0.5$,
$X_1 \eq (2\gamma_2 E_F)/(\gamma_1 \!+\! \gamma_2) - h_0$ and $X_2 \eq
(4\gamma_2 E_F)/(\gamma_1 \!+\! 4\gamma_2) - h_0$. Furthermore, $B \eq
1/(2\gamma_1 \!+\! 4\gamma_2)$ and
\begin{eqnarray}
  &D& = h_0(\gamma_1-2\gamma_2) +
  \nonumber \\
  & & 2\sqrt{  h_0^2(\gamma_1^2-\gamma_1 \gamma_2 -2\gamma_2^2)
    -2h_0 \gamma_2 E_F(\gamma_1-\gamma_2) +4 \gamma_2^2 E_F^2} \, .
  \nonumber
\end{eqnarray}
For Neel walls, $T_y(k_x,k_z)$ is not conformal under translation and
Eq.~\ref{Conductance:BW} does not apply.

Beyond the adiabatic approximation, we use a stable transfer matrix
method \cite{Usuki:prb95} to find the conductance numerically.  The
system is discretized into $N_x \times N_y \times N_z$ lattice points
with lattice constants $a_x$, $a_y$ and $a_z$. We use $L_z \eq L_x \eq
N_x a_x \eq 5$, $a_x \eq a_z \eq 0.05$, $L_y \eq 150$, $a_y \eq 0.08$
and $\gamma_1 \eq 6.8$. Varying $\gamma_2$ or $h_0$ changes the
reservoir properties.  We have chosen to vary the Fermi energy,
$E_F(\gamma_2,h_0)$, keeping the {\it hole carrier density constant},
to mimic experimental conditions.  $E_F(\gamma_2,h_0)$ is calculated
numerically.  We fixed the volume density of holes such that
$E_F(\gamma_2 \eq 2.1,h_0 \eq 0) \eq 1$.  Experimentally
\cite{Matsukura:prb98,Abolfath:prb01}, $E_F(h_0)/h_0 \!\sim\! 1$ which
corresponds to $h_0 \!\sim\! 1.5$.  The convergence of the numerical
scheme is verified by checking the unitarity of the scattering matrix.
We present our results in terms of relative domain wall resistance,
$\Re_{w} \eq (R-R_0)/R_0$.

\begin{figure}
   \includegraphics[bb=0 10 274 225, scale=0.9]{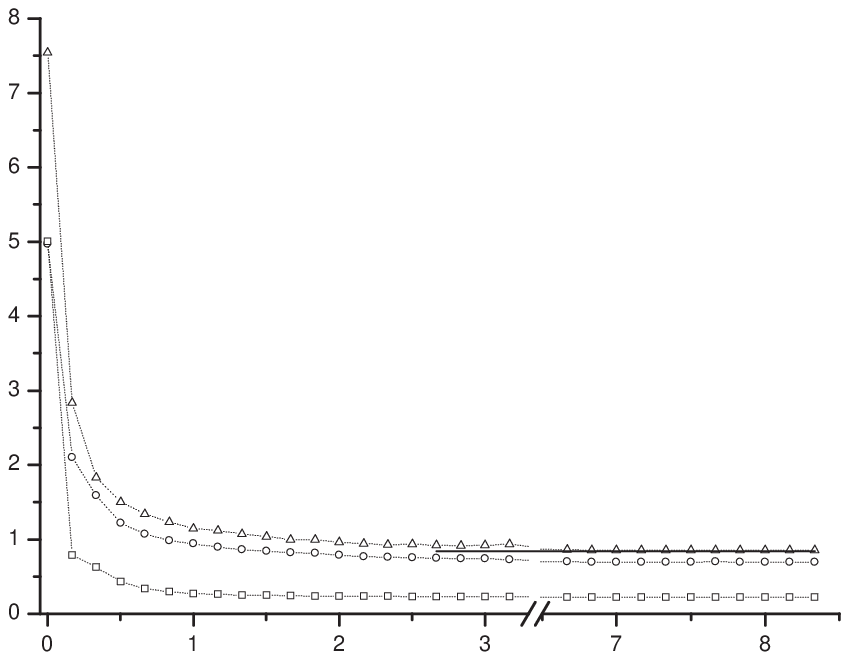}
   \put(-245,80){\rotatebox{90}{$\Re_{w}$}}
   \put(-120,0){$\lambda_{w}$}
   \put(-205,105){$h_0=1.5$}
   \put(-205,95){$\gamma_2=2.1$}
   \put(-211,120){\includegraphics[scale=0.25]{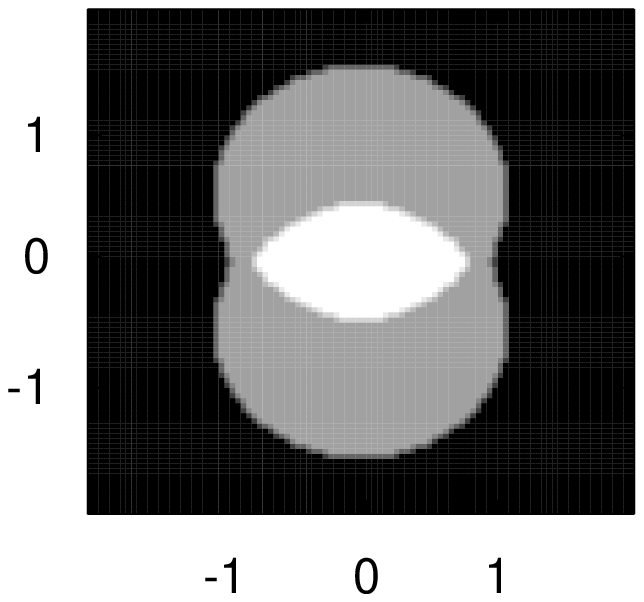}}
   \put(-169,120){\includegraphics[scale=0.25]{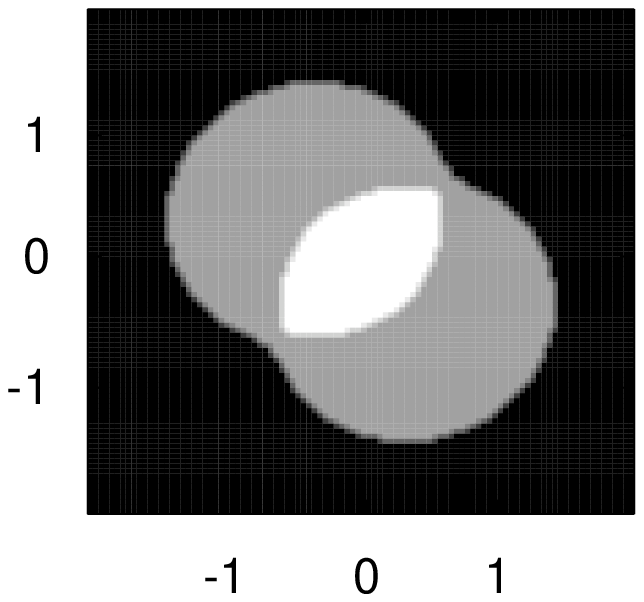}}
   \put(-127,120){\includegraphics[scale=0.25]{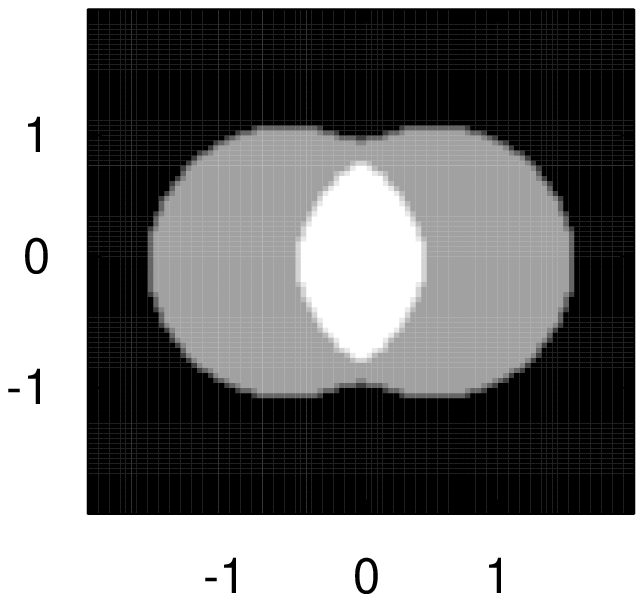}}
   \put(-85,120){\includegraphics[scale=0.25]{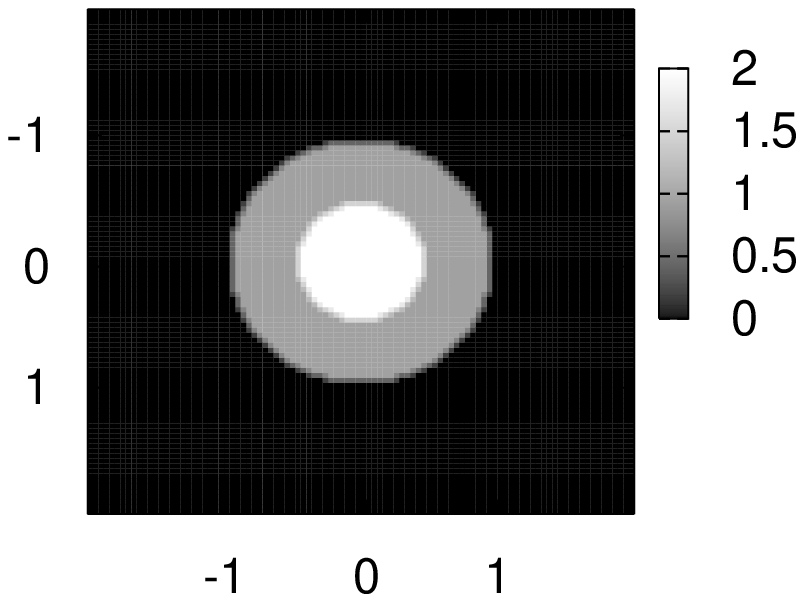}}
   \put(-59,123){$k_x$}
   \put(-101,123){$k_x$}
   \put(-143,123){$k_x$}
   \put(-184,123){$k_x$}
   \put(-215,150){\rotatebox{90}{$k_z$}}
   \put(-130,45){\includegraphics[scale=0.3]{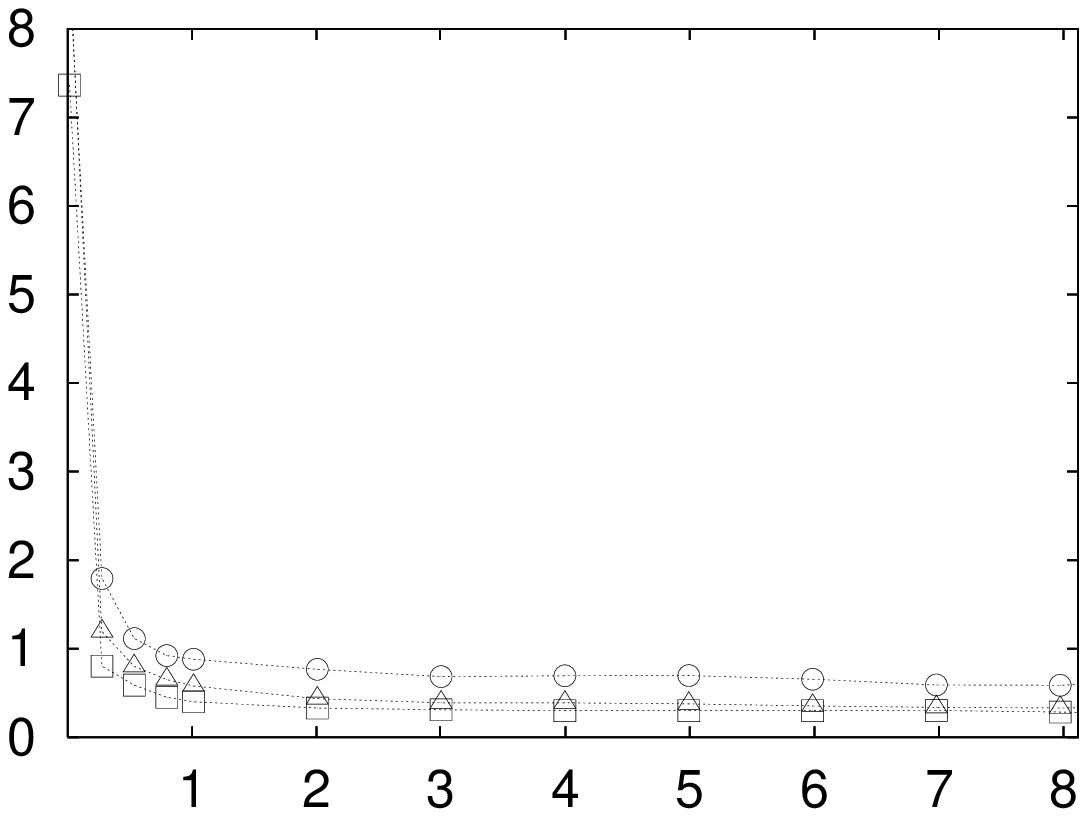}}
   \put(-135,75){\rotatebox{90}{$\Re_{w}$}}
   \put(-80,42){$\lambda_{w}$}
   \caption{Relative domain wall resistance versus wall width for ZX
     wall (triangle), ZY wall (circle), and YZ-wall (square). The
     solid line displays the intrinsic resistance derived from
     Eq.~\ref{Conductance:BW}.  Bottom right inset:, $\Re_{w}$ for a
     film shaped system with Dirichlet boundary conditions.  Top left
     to right insets: Distribution of conducting channels
     $T_y(k_x,k_z)$ for $y \eq -\infty$, $y \eq -0.88\lambda_{w}$, $y
     \eq 0$, and the intersection $T(k_x,k_z)$, for a ZX wall.}
\label{Fig1}
\end{figure}

For increasing $\lambda_{w}$, the direct collision between holes and
the domain wall gradient decreases and $\Re_w(\lambda_w)$ drops
rapidly to its intrinsic value, Fig.1. Note that
$\Re_{w}(\lambda_{w})$ is very close to its intrinsic value already
when $\lambda_{w} \!\sim\! 1$ for ZX and ZY walls, and $\lambda_{w}
\!\sim\!  0.5$ for a YZ wall {\em e.g.} when the domain wall width is
equal or smaller than the Fermi wavelength of heavy holes.

\begin{figure}
   \includegraphics[bb=0 10 274 225, scale=0.9]{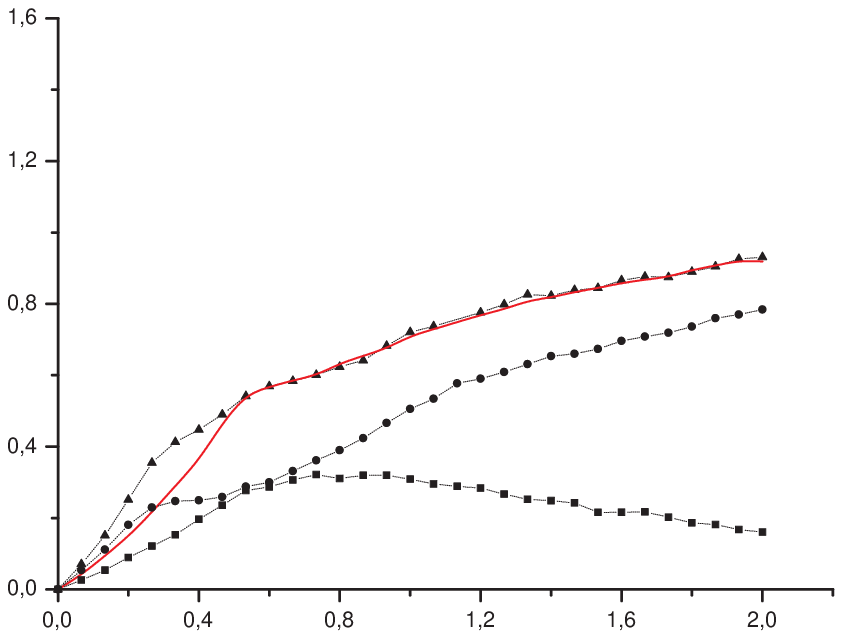}
   \put(-245,80){\rotatebox{90}{$\Re^I_{w}$}}
   \put(-120,0){$h_0$}
   \put(-65,155){$\gamma_2=2.1$}
   \put(-200,100){\includegraphics[scale=0.3]{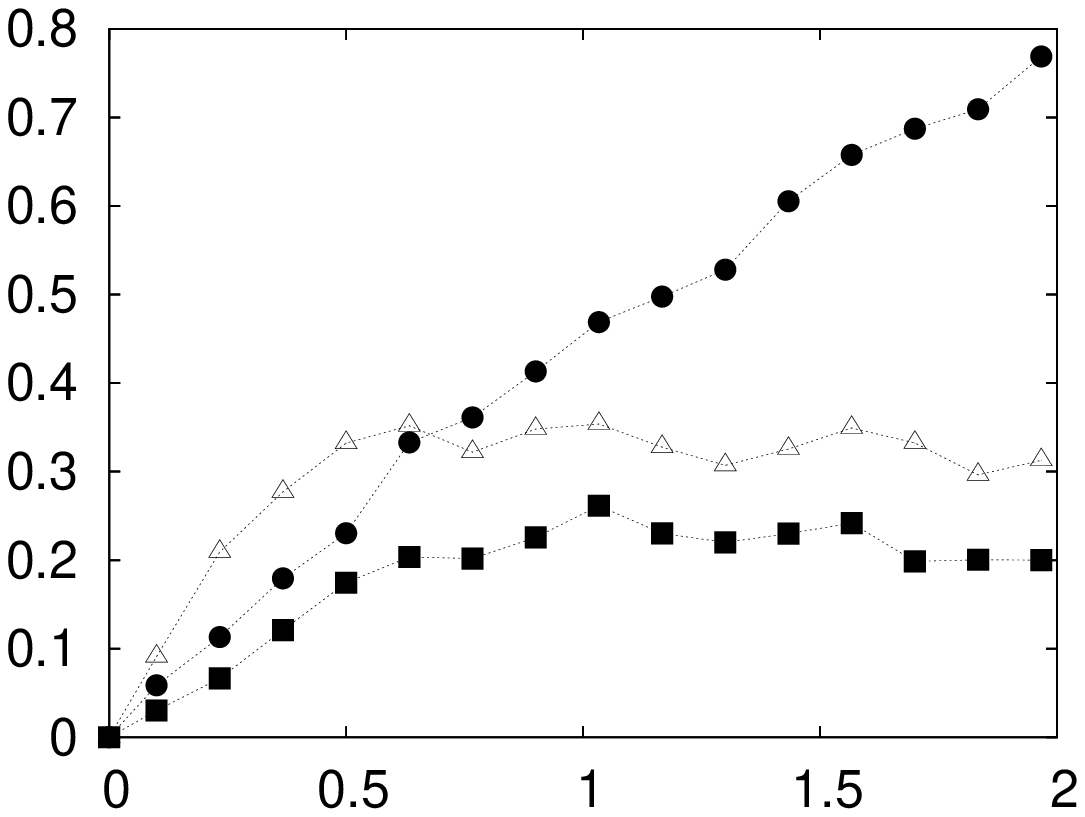}}
   \put(-205,130){\rotatebox{90}{$\Re^I_{w}$}}
   \put(-150,95){$h_0$}
   \caption{Intrinsic relative domain wall resistance versus exchange
     field for ZX-wall (triangle), ZY-wall (circle), and YZ-wall
     (square). The solid line displays the analytical result for a
     ZX-wall, Eq. (3).  Inset: $\Re^I_{w}(h_0)$ for a film shaped
     system with Dirichlet boundary conditions.}
\label{Fig2}
\end{figure}

The relative intrinsic domain wall resistance as a function of the
exchange field is shown in Fig.2. Generally, $\Re_{w}^I(h_0)$
increases for increasing $h_0$, due to an increase in the anisotropy
in $T_y(k_x,k_z)$.  For the YZ wall, however, the Sharvin resistance
$R_0$ increases even faster than $R$ leading to a weak reduction in
$\Re_{w}^I$ at large $h_0$. We also see in Fig.2 that the numerical
result for the Bloch wall is, within $2\%$, identical to the
analytical result derived from Eq.~\ref{Conductance:BW} for $h_0 >
0.5$.  There are small deviations at small $h_0 < 0.5$. Here, the
adiabatic limit requires very large $\lambda_{w}$ and system sizes
larger than one we used, $L_y \eq 150$, for convergence.

\begin{figure}
   \includegraphics[bb=0 10 274 225, scale=0.9]{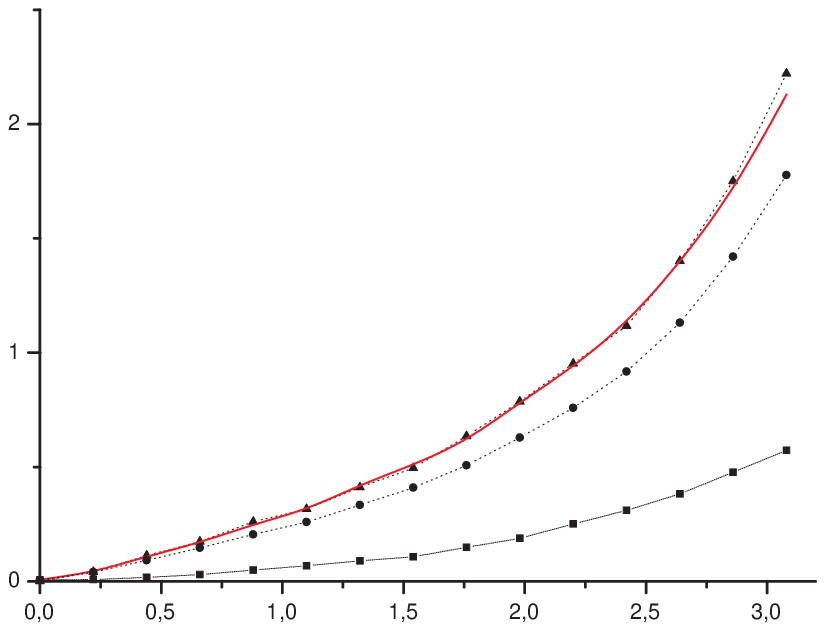}
   \put(-245,80){\rotatebox{90}{$\Re^I_{w}$}}
   \put(-120,0){$\gamma_2$}
   \put(-85,150){$h_0=1.5$}
   \put(-210,90){\includegraphics[scale=0.3]{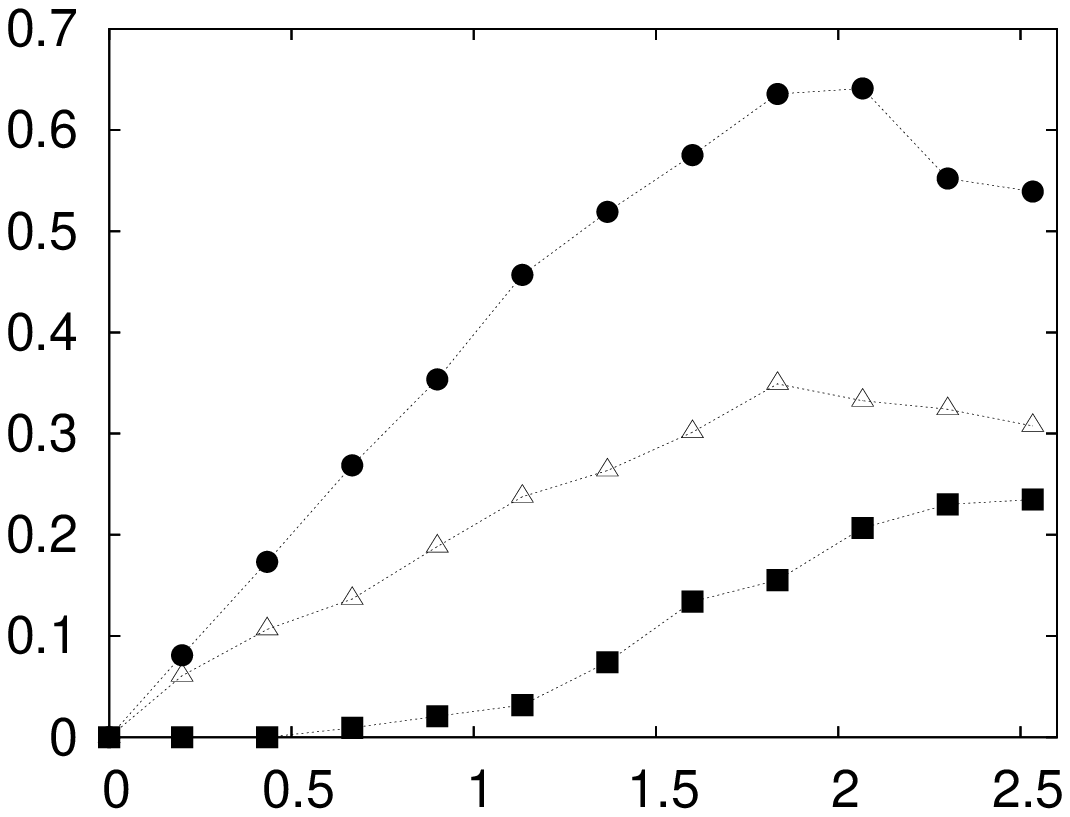}}
   \put(-215,120){\rotatebox{90}{$\Re^I_{w}$}}
   \put(-155,85){$\gamma_2$}
   \caption{Intrinsic relative domain-wall resistance as functions of
     hole spin-orbit coupling parameter, $\Re^I_{w}(\gamma_2)$:
     ZX-wall (triangle), ZY-wall (circle) and YZ-wall (square). Solid
     line shows the analytical result for a ZX-wall. Inset:
     $\Re^I_{w}(\gamma_2)$ for a film shaped system with Dirichlet
     boundary conditions.}
\label{Fig3}
\end{figure}

The relative intrinsic domain wall resistance as a function of the
effective spin-orbit coupling is shown in Fig. 3.
$\Re_{w}^I(\gamma_2)$ increases monotonically with increasing
$\gamma_2$ for all walls.  This is due to an increase in the
anisotropy in $T_y(k_x,k_z)$.  We also see that the numerical result
for the Bloch wall is identical to the analytical result derived from
Eq.~\ref{Conductance:BW} for $\gamma_2 < 2.7$. The small deviation at
large $\gamma_2 > 2.7$ is caused by the same effect as previously
discussed for $\Re^I_{w}(h_0<0.5)$. Here, increasing $\gamma_2$
reduces the effective amplitude of the exchange field along certain
directions.

For reasonable parameters $\gamma_2 \!\sim\! 2.1$ and $h_0 \!\sim\!
1.5$, $\Re^I_{w} \!\sim\! 0.9$ and $0.7$ for the ZX and $ZY$ walls,
respectively.  In other words, a domain wall removes nearly half of
the Sharvin conducting channels in ballistic adiabatic transport, an
effect that should be clearly measurable. An interesting question is
how much of the intrinsic resistance still remains in the diffusive
transport regime.  We know that the anisotropy in the distribution of
conducting channels with respect to the direction of the exchange
field do survive in the diffusive limit leading to the AMR effect
\cite{Hayashi:physicaB00,Baxter:prb02,Jungwirth:apl03,Wang:prb05}.  We
therefore expect that at least part of the intrinsic resistance will
also survive the diffusive limit. Furthermore, the intrinsic domain
wall resistance is reached already for domain wall widths comparable
to the Fermi wavelength. Consequently, we expect that the ballistic
intrinsic domain wall resistance will be important even in rather
dirty state-of-the art dilute ferromagnetic semiconductors where the
mean free paths are not very much smaller than the Fermi wavelengths.

Let us explain the rigidity of the intrinsic domain wall resistance
against variations in the domain wall's width and shape using
topological arguments.  Consider a Bloch ZX wall.  First, define the
order parameter space as a two dimensional space in which two
dimensional vectors map to points \cite{Chaikin:book95}, {\it e.g.}
the real space vector $(h_x \hat{x} \!+\! h_z \hat{z})$ maps into the
point $(M_x,M_z) \eq (h_x,h_z)$ in the order parameter space.  Second,
note that the distribution of conducting channels, $T_y(k_x,k_z)$,
depends on the average exchange field within a wave packet
$\mathbf{h}(y) \rightarrow \tilde{\mathbf{h}}(y)$.  Thus, the
conductance $G \!\sim\!  \cap_{y=-\infty}^{\infty} T_y(k_x,k_z)$
depends on $\tilde{\mathbf{h}}(y)$ or, more precisely, its mapping in
the order parameter space.  For $\lambda_{w} \ll \lambda_F$,
$\tilde{\mathbf{h}}(y)$ maps, approximately, into a half ellipse,
$M_x^2 \lambda_F^2/9\lambda_{w}^2 \!+\!  M_z^2 \eq h_0^2$ where
$M_x>0$. The mapped curve becomes more circular for increasing
$\lambda_w$.  For $\lambda_{w} > \lambda_F$, $\tilde{\mathbf{h}}(y)$
maps to a half circle, $M_x^2 \!+\!  M_z^2 \eq h_0^2$ where $M_x>0$.
In this smooth (adiabatic) limit, $ZX$ domain walls of all widths and
shapes map into the same half circle, and thus have the same
conductance.

So far, we have assumed periodic boundary condition in the transverse
directions. Spin and orbital motion are coupled at the boundaries due
to the spin-orbit interaction, which could change the results. In
order to address this question, we have also numerically computed the
conductance using Eq.~\ref{Hamiltonian} with Dirichlet boundary
condition, $\psi(0,y,z) \eq \psi(L_x \!+\!  a_x,y,z) \eq \psi(x,y,0)
\eq \psi(x,y,L_z \!+\! a_z) \eq 0$.  To mimic experimental available
3D film shaped conductors, we use $L_x \eq 6$, $a_x \eq 1/3$, $L_z \eq
3$, $a_z \eq 3/10$ and $L_y \eq 100$, $a_y \eq 1/4\pi$. It turns out
that the Dirichlet boundary condition together with small $L_z$
prevent the development of extreme anisotropies in the distribution of
conducting channels, expected at large $h_0$ and $\gamma_2$. This
leads to a reduction of $\Re_{w}^I$ by a factor of 2 for the ZX wall.
$\Re_{w}^I$ for the ZY and YZ walls are less affected since the
distribution of conducting channels for $\mathbf{h} \| \hat{y}$ is
isotropic.  The relative domain wall resistance as a function of
$\lambda_{w}$ is shown in the inset of Fig.~\ref{Fig1}.  Similar to
the PBC case, $\Re_{w}(\lambda)$ drops to its intrinsic value at
$\lambda_{w} \!\sim\! 1$ for $ZX$ and $ZY$ walls, and $\lambda_{w}
\!\sim\!  0.5$ for YZ wall.  $\Re_{w}^I(h_0)$ is shown in the inset of
Fig.2.  Here, $\Re_{w}^I$ for the ZX-wall forms a plateau for $h_0 >
0.5$ above which the anisotropy in the conducting channels is
prevented to develop further.  We see in the inset of Fig.~\ref{Fig3}
that $\Re_{w}^I(\gamma_2)$ for the ZY and ZX walls develop a
peak/plateau around $\gamma_2 \!\sim\! 2$ where the anisotropy in the
conducting channels is prevented to develop further.  In any case, the
intrinsic resistance remains well defined and finite in a 3D film
shaped conductor with Dirichlet boundary condition.

In conclusion, we have shown that dilute magnetic semiconductors
exhibits an intrinsic domain wall resistance. The intrinsic domain
wall resistance depends only on the map of the domain wall in the
magnetic order parameter space and not on their real space width and
shape. An analytical expression for the adiabatic conductance for
Bloch walls is given. For typical parameters, the intrinsic domain
wall resistance is comparable to the Sharvin resistance, and should
therefore be experimentally measurable. The domain wall resistance
drops to its intrinsic value when $\lambda_{w}$ approaches half the
Fermi wavelength for heavy holes or a quarter of $\lambda_F$ for light
holes. These values are not too far from mean free path accessible
presently accessible, {\it e.g.}  in Ref.~\cite{Ruster:prl03}. From
general topological arguments, we show that topological magnetic
defects have an intrinsic resistance against transport of carriers
with strong spin-orbit coupling that survive the adiabatic limit.

We thank J.\ Schliemann, Y.\ Galperin, A. Sudb{\o}, J.\ Hove, 
D.\ Huertas-Hernando 
and  G.\ E.\ W.\ Bauer for stimulating discussions. This work has been 
supported in part by the Research Council of Norway through grants no.\  
162742/V00, 1534581/432, 1585181/143, 
and 1585471/431.
   

\end{document}